\documentclass[aps,prl,twocolumn]{revtex4-2}
\usepackage{graphicx}
\usepackage[english]{babel}
\usepackage[T1]{fontenc}
\usepackage{mathrsfs}
\usepackage[tbtags]{amsmath}
\usepackage{amssymb}
\usepackage{amsxtra}
\usepackage{amsopn}
\usepackage{latexsym}
\usepackage[mathcal]{eucal}
\usepackage{mathtools}
\usepackage{slashed}
\usepackage{xcolor}
\usepackage[hyperfootnotes=false,bookmarks=false]{hyperref}

\usepackage{braket}
\usepackage{enumerate}
\usepackage{lipsum}

\newcommand{\BE}{\begin{equation}}
\newcommand{\EE}{\end{equation}}
\newcommand{\BA}{\begin{align}}
\newcommand{\EA}{\end{align}}

\newcommand{\mc}{\mathcal}
\newcommand{\psibar}{\overline{\psi}}
\newcommand{\cbar}{\overline{c}}

\begin{document}

\title{Nonperturbative Higgs-Schwinger mechanism at the origin\\of the gluon mass and color confinement}

\author{Giorgio Comitini}
\email{giorgio.comitini@dfa.unict.it}
\affiliation{Dipartimento di Fisica e Astronomia ``E. Majorana'', Universit\`a di Catania, Via S. Sofia 64, I-95123 Catania, Italy}

\date{\today}

\begin{abstract}
   Evidence from lattice and continuum studies supports the existence of a fully nonperturbative Higgs mechanism generating mass for gluons in linear covariant gauges. The broken charge is the Kugo-Ojima charge. The corresponding unphysical Goldstone boson is a bound-state superposition of two gluons, three gluons, a ghost-antighost, and a quark-antiquark pair. Mass generation occurs via the Schwinger mechanism, triggered by the formation of the Goldstone boson. Once corrected for symmetry breaking, the color charge operator is unbroken and confining.
\end{abstract}

\maketitle

\section{Introduction}

In the Standard Model of particle physics, the elementary degrees of freedom that make up visible matter~-- quarks, leptons and gauge bosons -- are described as massless at their most fundamental level. The discovery of the Higgs boson at the Large Hadron Collider in 2012 \cite{HiggsAtlas,HiggsCMS} confirmed the Brout-Englert-Higgs (BEH) scenario of mass generation \cite{EB64,Higgs64,GHK64}, by which quarks, charged leptons, the gauge bosons of the electroweak sector \cite{Gla61} -- except for the photon -- and the Higgs boson itself acquire a mass through the vacuum condensation of a Higgs field and the subsequent breaking of the SU(2)$_{L}\times$U(1)$_{Y}$ electroweak symmetry down to the electromagnetic U(1)$_{em}$ \cite{Wei67,Sal68}. The BEH theory, however, explains but a tiny fraction of the mass of the visible universe. More than 98\% of said mass, in fact, comes from nuclei, and is generated by the strong interactions of quarks and gluons \cite{RS20}.

\begin{figure}[h]
    \includegraphics[width=0.38\textwidth]{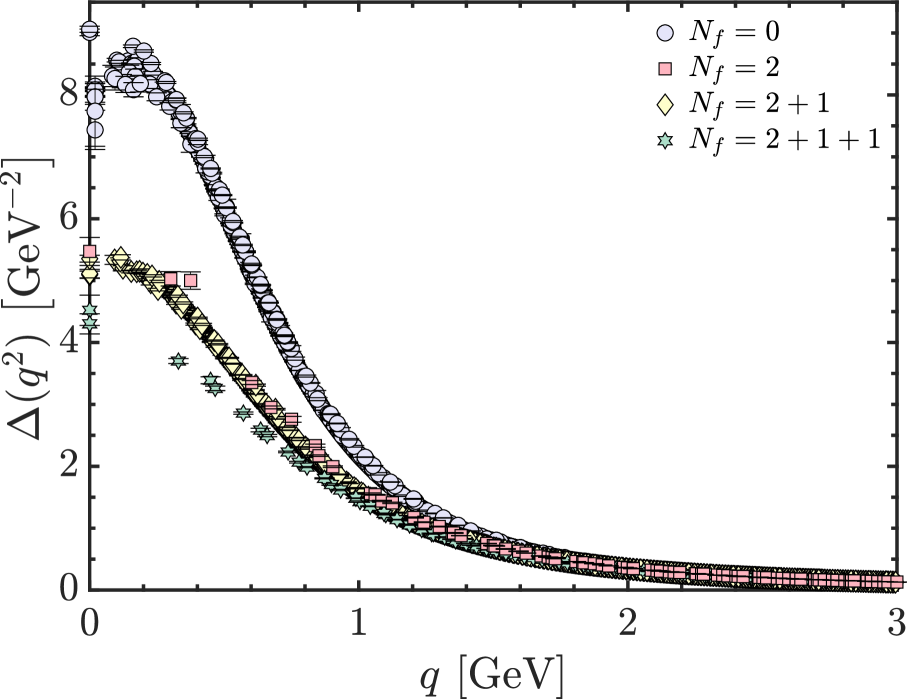}
    \caption{Lattice Landau-gauge Euclidean transverse gluon propagator as a function of momentum for different quark numbers $N_{f}$. Figure from \cite{FP25}, data from \cite{BIMS09,ABBC12,BRR17,DOS18,ADFP20,AADF21}.}
    \label{gluprop}
\end{figure}

At the turn of the century, simulations performed on huge lattices \cite{LSWP98a,LSWP98b,BBLW00,BBLW01,BHLP04,SIMS05,CM08b,BIMS09,ISI09,BMM10,ABBC12,BLLM12,OS12,BBCO15} lead to a groundbreaking discovery: instead of growing to infinity as is typical for a massless particle, the transverse component of the gluon propagator of linear covariant gauges saturates to a finite, nonzero value in the limit of vanishing momentum -- see Fig.~\ref{gluprop}. This property entails that no massless transverse gluons exist either in the physical or in the unphysical spectrum of QCD, and is thus often summarized in the statement that, in the infrared regime of the strong interactions, gluons acquire a dynamically generated mass \cite{Cor81}. The gluon mass protects QCD from developing infrared instabilities such as the Landau pole that affects the running of its perturbative coupling \cite{SIMS05,BIMS09,ABBC12,BLLM12}; by doing so, it sets the scale for the low-energy dynamics and phenomenology of the strong interactions.

Despite the overwhelming consensus surrounding gluon dynamical mass generation, its underlying mechanism remains a matter of intense debate. On the one hand, the restriction of the Euclidean path integral of Landau-gauge QCD to the so-called first Gribov region -- required to (partially) lift the Gribov-Singer ambiguity \cite{Gri78,Sin78} that affects non-abelian gauge fixing -- is known to yield an infrared-finite gluon propagator \cite{Zwa89,DGSVV08,TW11,PTW13,CFGM15,RSTW17}. On the other hand, the vector bosons of gauge theories can acquire a mass regardless of the non-abelian nature of the gauge group -- and in particular of the Gribov-Singer ambiguity -- through the so-called Schwinger mechanism \cite{Sch62a,Sch62b,EF74,Smi74,PTT75,Cor81,AP06,AIMP12,IP13,ABFP18,FP25}. When the latter is in action, mass generation is triggered by the formation of massless poles in the interaction vertices of the theory and the subsequent evasion of the identity that protects gauge bosons from becoming massive: the seagull identity \cite{FP25}. Strong evidence in favor of the Schwinger mechanism being active in QCD was obtained very recently \cite{AFP22,ADF23,AFIP23,FP24}.

The purpose of this Letter is to show that evidence from lattice and continuum studies supports the existence of a fully nonperturbative Higgs mechanism acting in the gauge sector of QCD in linear covariant gauges. There are three ingredients to any Higgs mechanism: a broken charge, a corresponding unphysical Goldstone boson ``eaten'' by the gauge boson, and a concrete mechanism by which mass is generated. In QCD, the Schwinger mechanism provides the latter. The broken charge is the Kugo-Ojima charge~-- that is, the part of the Noether color charge which is exact under Becchi-Rouet-Stora-Tyutin (BRST) transformations ~--, and the Goldstone boson can be interpreted as a massless colored bound-state superposition of two gluons, three gluons, a ghost-antighost, and a quark-antiquark pair. The Goldstone boson triggers the Schwinger mechanism responsible for gluon mass generation. Due to symmetry breaking, the color charge operator fails to generate color rotations of the Goldstone boson. However, the operator can be redefined so that color symmetry is preserved unbroken on the asymptotic states -- that is, on scattering states. The resulting color charge is exact under BRST transformations, which implies that its matrix elements between physical states vanish: color is confined.

Since the mechanism is also active in the absence of quarks, in what follows we will mostly frame our discussion in the context of pure Yang-Mills theory.

\newpage

\section{A Goldstone boson for the massive gluon}

Pure Yang-Mills theory in a generic linear covariant gauge is defined by the Faddeev-Popov action
\begin{align}
   \mc{S}_{\text{FP}}&=\int d^{4}x\ \Big\{-\frac{1}{4}\,F_{\mu\nu}^{a}F^{a\,\mu\nu}-i\partial^{\mu}\cbar^{a}D_{\mu}c^{a}+\\
   \notag&\qquad\qquad\qquad\quad \ +\frac{\xi}{2}\,B^{a}B^{a}+B^{a}\partial^{\mu}A_{\mu}^{a}\Big\}\ ,
\end{align}
where $A_{\mu}^{a}$ is the gluon field, $F_{\mu\nu}^{a}=\partial_{\mu}A_{\nu}^{a}-\partial_{\nu}A_{\mu}^{a}+gf^{abc}A_{\mu}^{b}A_{\nu}^{c}$ is the field-strength tensor, $c^{a}$ and $\cbar^{a}$ are the Faddeev-Popov ghost and antighost fields, $B^{a}$ is the Nakanishi-Lautrup auxiliary field -- satisfying the constraint equation $\partial^{\mu}A_{\mu}^{a}+\xi B^{a}=0$ --, $D_{\mu}^{ab}=\delta^{ab}\partial_{\mu}-gf^{abc}A_{\mu}^{c}$ is the covariant derivative in the adjoint representation, $f^{abc}$ are the structure constants of the gauge group and $\xi$ is the gauge parameter. The action $\mc{S}_{\text{FP}}$ is invariant under the transformation
\begin{align}\label{brst}
   \delta_{B}A_{\mu}^{a}&=D_{\mu}c^{a}\ ,\quad\delta_{B}c^{a}=-\frac{g}{2}f^{abc}c^{b}c^{c}\ ,\\
   \notag \delta_{B} \cbar^{a}&=iB^{a}\ ,\qquad \delta_{B}B^{a}=0\ ,
\end{align}
generated by a BRST charge operator $Q_{B}$ that acts on a generic field $\mc{F}$ as $\delta_{B}\mc{F}=[iQ_{B},\mc{F}]_{\mp}$, where $[\cdot,\cdot]_{\mp}$ is a commutator or an anticommutator, depending on the nature of $\mc{F}$.

BRST symmetry reflects the residual gauge invariance of the gauge-fixed action $\mc{S}_{\text{FP}}$. Because of it, the momentum-space gluon propagator $\Delta_{\mu\nu}^{ab}(p)=\int d^{4}x\ e^{ipx}\, \langle T\left\{A_{\mu}^{a}(x)A_{\nu}^{b}(0)\right\}\rangle$ has the form \footnote{In what follows we will frequently suppress diagonal color indices.}
\begin{align}\label{gprop}
   \notag\Delta_{\mu\nu}(p)&=\int dm^{2}\ \sigma(m^{2})\ \left(-\eta_{\mu\nu}+\frac{p_{\mu}p_{\nu}}{m^{2}}\right)\ \frac{i}{p^{2}-m^{2}}+\\
   &\qquad-\frac{p_{\mu}p_{\nu}}{M^{2}}\,\frac{i}{p^{2}}+\xi p_{\mu}p_{\nu}\,\frac{-i}{(p^{2})^{2}}\ ,
\end{align}
where $\sigma(m^{2})$ is the spectral function associated to the transverse component $\Delta(p^{2})$ of the propagator \footnote{In the presence of complex-conjugate singularities, one must use a suitably generalized spectral representation.},
\begin{equation}
   \Delta(p^{2})=\frac{1}{3}\,t^{\mu\nu}(p)\Delta_{\mu\nu}(p)=-i\int dm^{2}\ \frac{\sigma(m^{2})}{p^{2}-m^{2}}\ ,
\end{equation}
$t_{\mu\nu}(p)=\eta_{\mu\nu}-p_{\mu}p_{\nu}/p^{2}$ being the transverse projector. The constant $M^{-2}=-i\Delta(0)=\int dm^{2}\ \sigma(m^{2})/m^{2}$ encodes the zero-momentum behavior of the gluon propagator; it is known to be finite and nonvanishing thanks to both lattice simulations \cite{LSWP98a,LSWP98b,BBLW00,BBLW01,BHLP04,SIMS05,CM08b,BIMS09,ISI09,BMM10,ABBC12,BLLM12,OS12,BBCO15} and continuum studies \cite{AvS01,AN04,AP06,ABP08,FMP09,HvS13} -- see Fig.~\ref{gluprop}. This property of QCD is summarized by the statement that, in the infrared, gluons acquire a dynamically generated mass.

The first term in Eq.~\eqref{gprop} describes the massive component of the gluon spectrum. While no consensus currently exists on its content \cite{CDMV12,DOS14,Sir16b,HK19,BT20,LLOS20}, the presence of the massive free vector propagator $-\eta_{\mu\nu}+p_{\mu}p_{\nu}/m^{2}$ shows that it comprises spin-1 states arranged in transverse polarization triplets. The second and third terms in Eq.~\eqref{gprop}, on the other hand, contain scalar poles at $p^{2}=0$, and fully determine the massless component of the gluon spectrum. In modern studies of the gluon mass, this component is usually ignored.

The nature of the massless states in the gluon spectrum can be understood by employing the techniques of asymptotic analysis developed in \cite{KO79b,KO79c}. The analysis reveals that the massless poles in the gluon propagator of Eq.~\eqref{gprop} correspond to two asymptotic scalar fields $\chi^{a}$ and $\beta^{a}$ which -- modulo renormalization -- appear in the field operators $A_{\mu}^{a}$ and $B^{a}$ as
\begin{equation}\label{asym}
   A_{\mu}^{a}\to M^{-1}\partial_{\mu}\chi^{a}+\cdots\ ,\qquad B^{a}\to M \beta^{a}+\cdots\ .
\end{equation}
Here the arrows indicate that the asymptotic limit $t\to \pm \infty$ has been taken and the ellipses denote unspecified massive asymptotic fields. $\chi^{a}$ and $\beta^{a}$ create and annihilate massless in and out states in the asymptotic Fock space of the theory. They satisfy the field equations $\partial^{2}\chi^{a}=-\xi M^{2}\beta^{a}$, $\partial^{2}\beta^{a}=0$, and have commutators
\begin{align}\label{commsp}
   \notag[\chi^{a}(x),\chi^{b}(y)]&=-i\delta^{ab}\left\{D(x-y)-\xi M^{2}E(x-y)\right\}\ ,\\
   [\chi^{a}(x),\beta^{b}(y)]&=-i\delta^{ab}\,D(x-y)\ ,\quad[\beta^{a}(x),\beta^{b}(y)]=0\ ,
\end{align}
where $iD(x-y)$ is the commutator of a free massless scalar field and $E(x)$ is the dipole Green function satisfying $\partial^{2}E(x)=D(x)$. In the Landau gauge ($\xi=0$), extra machinery required for the dipole can be avoided, and to Eq.~\eqref{commsp} correspond
\begin{equation}\label{commca}
   [\chi^{a}_{\bf p},\chi^{b\,\dagger}_{\bf q}]=[\chi^{a}_{\bf p},\beta^{b\,\dagger}_{\bf q}]=-\delta^{ab}\,(2\pi)^{3}\,\delta^{(3)}({\bf p}-{\bf q})
\end{equation}
as the only nonvanishing commutators between the creation and annihilation operators of $\chi^{a}$ and $\beta^{a}$. From Eq.~\eqref{commca} it follows that $\chi^{a}$ creates negative-norm states, whereas $\beta^{a}$ creates zero-norm states: in both cases, their quanta belong to the \textit{unphysical} sector of the spectrum of QCD. Under BRST, $\chi^{a}$ and $\beta^{a}$ transform as
\begin{align}\label{brste}
   [i Q_{B},\chi^{a}]&=M\gamma^{a}\ ,\qquad\{iQ_{B},\gamma^{a}\}=0\ ,\\
   \notag\{iQ_{B},\overline{\gamma}^{a}\}&=iM\beta^{a}\ ,\qquad[iQ_{B},\beta^{a}]=0\ ,
\end{align}
where $\gamma^{a}$ and $\overline{\gamma}^{a}$ -- with anticommutation relations
\begin{equation}
   \{\gamma^{a}_{\bf p},\overline{\gamma}^{b\,\dagger}_{\bf q}\}=i\delta^{ab}\,(2\pi)^{3}\,\delta^{(3)}({\bf p}-{\bf q})\ -
\end{equation}
are the asymptotic fields associated to the ghost and antighost fields $c^{a}$ and $\cbar^{a}$. In jargon, one says that, together with $\gamma^{a}$ and $\overline{\gamma}^{a}$, $\chi^{a}$ and $\beta^{a}$ form the elementary BRST quartet of QCD, of which $\chi^{a}$ is the so-called first parent \cite{Kug78,KO79a}. The BRST condition for physical states $Q_{B}\ket{\text{phys}}=0$ ensures that particles from the quartet always appear in physical states as zero-norm combinations, which decouple from the physical dynamics of the theory \cite{KO78b}.

Crucially, the asymptotic structure we just described is identical to that of theories in which mass generation occurs via a Higgs mechanism that preserves the symmetry between the gauge bosons' masses. For instance, in the SU(2) Higgs-Kibble model \footnote{Quantized in linear covariant gauges with massless ghosts and Goldstone bosons, not in $R_{\xi}$ gauges.} \cite{tHoo71}, the asymptotic gauge fields read \cite{KO79b}
\begin{equation}
   A_{\mu}^{i}\to U_{\mu}^{i}+M^{-1}\partial_{\mu}\chi^{i}\ ,\qquad B^{i}\to M\beta^{i}\ ,
\end{equation}
where $U_{\mu}^{i}$ is the asymptotic field of the massive gauge boson, $\chi^{i}=\lambda^{i}+(1-\xi N M)\beta^{i}$ is a linear combination of the unphysical components of the Higgs field $\lambda^{i}$ and of the Nakanishi-Lautrup field, and $N$ is a constant. $\chi^{i}$ and $\beta^{i}$ satisfy the same commutation relations and field equations as $\chi^{a}$ and $\beta^{a}$ and are part of the elementary BRST quartet of the SU(2) Higgs-Kibble model. Both in QCD with gluon dynamical mass generation and in the SU(2) Higgs-Kibble model, since $\partial^{\mu}A_{\mu}+\xi B=0$, $\beta$ is the massless scalar polarization of the gauge boson. In the SU(2) Higgs-Kibble model, modulo mixing with $\beta^{i}$, the $\chi^{i}$'s are the Higgs field's Goldstone bosons, eaten by the gauge bosons in the process of the latter acquiring a mass. What are the $\chi^{a}$'s in QCD? The answer to this question lies within the Yang-Mills field equations.

In the framework of BRST quantization, the operatorial Maxwell-Yang-Mills equation reads
\begin{equation}
   \partial_{\mu}F^{a\,\mu\nu}+gf^{abc}A_{\mu}^{b}F^{c\mu\nu}-\partial^{\nu}B^{a}+igf^{abc}\partial^{\nu} \cbar^{b}c^{c}=0\ .
\end{equation}
The equation can be evaluated between the vacuum state $\ket{\Omega}$ and the asymptotic state $\ket{\chi^{a}(p)}$ containing a single boson $\chi^{a}$ of momentum $p$. Since $\chi^{a}$ appears in $A_{\mu}^{a}$ as a gradient, the antisymmetric field-strength tensor $F_{\mu\nu}^{a}$ has no massless component: $\bra{\Omega}F_{\mu\nu}^{a}(x)\ket{\chi^{d}(p)}=0$ \cite{KO79a}. Moreover, from Eqs.~\eqref{asym} and \eqref{commsp}, $\bra{\Omega}\partial^{\nu}B^{a}(x)\ket{\chi^{d}(p)}=iMp^{\nu}\,\delta^{ad}\, e^{-ipx}$. It follows that
\begin{equation}\label{BSS}
   gf^{abc}\bra{\Omega}A_{\mu}^{b}F^{c\mu\nu}+i\partial^{\nu} \cbar^{b}c^{c}\ket{\chi^{d}(p)}=iMp^{\nu}\,\delta^{ad}\, e^{-ipx}\ .
\end{equation}

\begin{figure}[h]
    \includegraphics[width=0.44\textwidth]{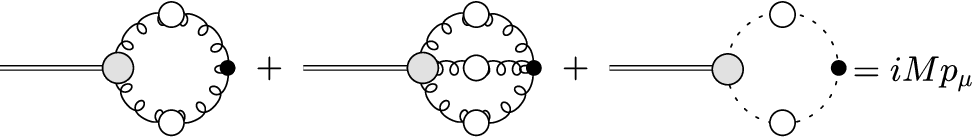}
    \caption{Diagrammatic depiction of the sum rule in Eq.~\eqref{BSSE}. The double line, curly line, dotted line, grey blobs and solid dots represent, respectively, the external Goldstone boson, the gluon propagator, the ghost propagator, the Goldstone boson's proper vertices and the tree-level vertices reported in the equation.}
    \label{sumrule}
\end{figure}

Eq.~\eqref{BSS} can be interpreted as a sum rule for so-called Bethe-Salpeter amplitudes \cite{BS51,RW94,AvS01} in the two-gluon, three-gluon and ghost-antighost channels -- respectively,
\begin{align}\label{BSA}
   B_{\mu\nu}^{abd}(x,y;p)&=\bra{\Omega}T\left\{A_{\mu}^{a}(x)A_{\nu}^{b}(y)\right\}\ket{\chi^{d}(p)}\ ,\\
   \notag B_{\mu\nu\sigma}^{abcd}(x,y,z;p)&=\bra{\Omega}T\{A_{\mu}^{a}(x)A_{\nu}^{b}(y)A_{\sigma}^{c}(z)\}\ket{\chi^{d}(p)}\ ,\\
   \notag B^{abd}(x,y;p)&=\bra{\Omega}T\{c^{a}(x)\cbar^{b}(y)\}\ket{\chi^{d}(p)}\ .
\end{align}
Roughly speaking, these amplitudes describe the wavefunction of the boson $\chi^{a}$ in terms of gluon and ghost constituents. Eq.~\eqref{BSS} -- see Fig.~\ref{sumrule} -- translates to
\begin{align}\label{BSSE}
   \notag&gf^{abc}\,\Big[\partial^{[\mu}_{(y)}B^{bcd\,\nu]}_{\mu}(x,y;p)+gf^{cef}B^{befd\,\mu\nu}_{\mu}(x,y,z;p)+\\
   &\qquad\quad +i\partial^{\nu}_{(y)}B^{bcd}(x,y;p)\Big]=iMp^{\nu}\,\delta^{ad}\, e^{-ipx}
\end{align}
in the limit $z\to y\to x$, and implies that at least one of the Bethe-Salpeter amplitudes in Eq.~\eqref{BSA} must be different from zero. In the absence of symmetries forcing their vanishing, it is reasonable to assume that all three of them will be. It follows that the boson $\chi^{a}$ can be interpreted as a massless scalar colored bound-state superposition of two gluons, three gluons, and a ghost-antighost pair. In the presence of quarks, Eq.~\eqref{BSS} also contains the quark color current $g\sum_{f} \psibar_{f}\gamma^{\nu}T^{a}\psi_{f}$, so $\chi^{a}$ can acquire a component in the quark-antiquark channel.

\section{Symmetry breaking, symmetry preservation and color confinement}

In the framework of BRST quantization, it is well-known that gauge bosons can only acquire a mass if either of two conditions are met: either the so-called Kugo-Ojima criterion is satisfied, or the global charge associated to the local gauge symmetry is broken \cite{KO79a}.

The Kugo-Ojima criterion was originally formulated and is often quoted as a criterion for color confinement. At its core, however, it is a criterion for the unbrokennes of the Kugo-Ojima charge $N^{a}$,
\begin{equation}
   N^{a}=\int d^{3}x\ \mc{N}_{0}^{a}\ ,\quad \mc{N}_{\mu}^{a}=\{Q_{B},D_{\mu}\cbar^{a}\}\ ,\quad\partial^{\mu}\mc{N}_{\mu}^{a}=0\ .
\end{equation}
In terms of the first parent $\chi^{a}$ of the elementary BRST quartet \footnote{In our conventions, the measure of integration of three-dimensional momentum-space distributions is $(2p^{0})^{-1}\frac{d^{3}p}{(2\pi)^{3}}$, so $\,(ip^{0})\,(2\pi)^{3}\,\delta^{(3)}({\bf p})$ does not vanish.},
\begin{align}
   &\bra{\Omega}N^{a}\ket{\chi^{b}(p)}=(1+u)\,(iMp^{0})\,(2\pi)^{3}\,\delta^{(3)}({\bf p})\,\delta^{ab}\ ,
\end{align}
the criterion reads $1+u=0$ \cite{KO79a}. If $1+u\neq 0$, then the Kugo-Ojima charge is spontaneously broken and the first parent of the elementary BRST quartet is the Goldstone boson associated to its breaking.

Thanks to lattice simulations \cite{NF00,FN03,ABP09,ABFP24}, today we know that, in four-dimensional pure Yang-Mills SU(3) theory, $1+u\neq 0$. It follows that the first parent of the elementary BRST quartet $\chi^{a}$ is the Goldstone boson associated to the breaking of the Kugo-Ojima charge. As we saw before, this boson is the massless bound state that appears in the gluon spectrum when the gluon acquires a mass.

The breaking of the Kugo-Ojima charge has been traditionally interpreted as a pathological feature of Yang-Mills theory with dynamical mass generation, because it leads to the breaking of the color charge operator $Q^{a}$ obtained from Noether's first theorem, $Q^{a}=G^{a}+N^{a}$ \cite{KO79a}, where
\begin{equation}
   G^{a}=\int d^{3}x\ \mc{G}_{0}^{a}\ ,\quad \mc{G}_{\mu}^{a}=\partial^{\nu}F_{\mu\nu}^{a}\ ,\quad\partial^{\mu}\mc{G}_{\mu}^{a}=0\ .
\end{equation}
When the gauge bosons acquire a mass, the charge $G^{a}$ -- being the surface integral of a chromoelectric field operator $F_{0i}^{a}=\mc{E}^{a}_{i}$ with no massless modes --  vanishes \cite{KO79a}, so that $Q^{a}=N^{a}$: if $N^{a}$ is spontaneously broken, then so is the color charge $Q^{a}$.

However, it should be noted that, in the presence of symmetry breaking, $Q^{a}$ cannot be considered the true generator of color rotations. Indeed, it is easy to verify \footnote{For instance, one can insert the identity operator $1=\ket{\beta}\bra{\beta}-\ket{\chi}\bra{\beta}-\ket{\beta}\bra{\chi}+\cdots$ -- the ellipsis denoting irrelevant projectors -- in the commutators and compute the latter between the vacuum and $\ket{\chi}$ without assuming anything about the transformation properties of $\chi^{a}$, only that $\beta^{a}$ transforms as expected.} that $\bra{\Omega}\chi^{a}(x)\ket{\Omega}=0$ and $\bra{\Omega}Q^{a}\ket{\chi^{b}(p)}=\bra{\Omega}N^{a}\ket{\chi^{b}(p)}\propto(1+u)\neq 0$ imply
\begin{equation}
   [Q^{a},\chi^{b}(x)]\neq igf^{abc}\chi^{c}(x)\ ,\quad [Q^{a},Q^{b}]\neq igf^{abc}Q^{c}\ .
\end{equation}
The failure of these identities originates in the very violation of the Kugo-Ojima criterion and expresses the fact that~-- in the presence of symmetry breaking -- the Noether color charge is unable to generate color rotations of the Goldstone boson.

Fortunately for the consistency of QCD, this failure can be cured by a redefinition of the color charge operator \footnote{Such redefinitions are not at all uncommon, regardless of mass generation. For instance, in the framework of the BRST quantization of ordinary QED, the electric charge operator too must be redefined additively because of the very existence of massless photonic modes \cite{KO79a}.}. Indeed, if we introduce an operator $C^{a}$
\begin{equation}
   C^{a}=\int d^{3}x\ \mc{C}_{0}^{a}\ ,\quad \mc{C}_{\mu}^{a}=(1+u)\,M\,\partial_{\mu}\beta^{a}\ ,\quad\partial^{\mu}\mc{C}_{\mu}^{a}=0\ ,
\end{equation}
where the last equality reflects the masslessness of the $\beta^{a}$ quanta, it is easy to verify that the operator $\hat{Q}^{a}$ defined as
\begin{equation}
   \hat{Q}^{a}=Q^{a}-C^{a}
\end{equation}
is unbroken: $\bra{\Omega}\hat{Q}^{a}\ket{\chi^{b}(p)}=0$. The subtraction of $C^{a}$ exactly cancels the source of symmetry breaking in $Q^{a}$, yielding a new conserved charge which on every asymptotic field $X$ -- except for $\chi^{a}$ -- acts just like the color charge:
\begin{equation}\label{ccom}
   [\hat{Q}^{a},X(x)]=[Q^{a},X(x)]\ .
\end{equation}
The latter holds for massive asymptotic fields because they commute with massless ones, and for $\beta^{b}$, $\gamma^{b}$ and $\overline{\gamma}^{b}$ because of their known commutativity with $\beta^{a}$. Since no massless fields with the quantum numbers of the elementary BRST quartet are known except for its members, and since asymptotic fields with different quantum numbers commute, only $\chi^{a}$ escapes Eq.~\eqref{ccom}. The way this happens is exactly as required to restore the color algebra on Goldstone states. For instance,
\begin{equation}
   \bra{\Omega}[Q^{a},Q^{b}]\ket{\chi^{c}(p)}=igf^{abd}\bra{\Omega}(Q^{d}+C^{d})\ket{\chi^{c}(p)}\ ,
\end{equation}
which follows from $[Q^{a},\beta^{b}]=ig f^{abc}\beta^{c}$ and from the spontaneous breaking of $Q^{a}$, becomes
\begin{equation}
   \bra{\Omega}[\hat{Q}^{a},\hat{Q}^{b}]\ket{\chi^{c}(p)}=igf^{abd}\bra{\Omega}\hat{Q}^{d}\ket{\chi^{c}(p)}\ ,
\end{equation}
and similarly for $\bra{\Omega}[\hat{Q}^{a},\chi^{b}]\ket{\Omega}= igf^{abc}\bra{\Omega}\chi^{c}\ket{\Omega}$.

Besides being unbroken, the redefined operator $\hat{Q}^{a}$ has two essential properties making it a good color charge: 1. it commutes with the $S$ matrix, $(\hat{Q}^{a})^{\text{out}}S=S(\hat{Q}^{a})^{\text{in}}$; and 2. it is BRST exact. Indeed, $C^{a}$ can be expressed as
\begin{equation}
   C^{a}=(1+u)\,\left\{Q_{B},\int d^{3}x\ \partial_{0}\overline{\gamma}^{a}\right\}\ ,
\end{equation}
from which $\hat{Q}^{a}=\{Q_{B},\cdot\}$ follows. The first property implies that color is conserved, whereas the second implies that, for every pair of physical states $\ket{\text{phys}}$ and $\ket{\text{phys}^{\prime}}$,
\begin{equation}
   \bra{\text{phys}}\hat{Q}^{a}\ket{\text{phys}^{\prime}}=0\ :
\end{equation}
the color carried by the operator $\hat{Q}^{a}$ is confined.

\section{Discussion}

In this Letter we showed that evidence obtained from lattice and continuum studies of QCD -- namely, the saturation of the gluon propagator to a finite nonzero value \cite{LSWP98a,LSWP98b,BBLW00,BBLW01,BHLP04,SIMS05,CM08b,BIMS09,ISI09,BMM10,ABBC12,BLLM12,OS12,BBCO15,AvS01,AN04,AP06,ABP08,FMP09,HvS13} and the failure of the Kugo-Ojima criterion \cite{NF00,FN03,ABP09,ABFP24}~--, when framed in the context of the BRST quantization of QCD, supports the existence of a Higgs mechanism generating mass in the gauge sector of the strong interactions. Said mechanism is fully nonperturbative, in that the Goldstone bosons associated to symmetry breaking are not the quanta of an independent Higgs field, but rather colored bound-state superpositions of two gluons, three gluons, a ghost-antighost, and a quark-antiquark pair. While the spontaneous breaking of the Kugo-Ojima charge implies the breaking of the Noether color charge, a color charge operator can still be defined which is unbroken, conserved, BRST exact and thus confining.

Within the picture presented in this Letter, gluon mass generation and color confinement take place as follows. First, the dynamics of QCD is such that it allows the formation of a massless bound state with the quantum numbers of the scalar gluon -- that is, of the color charge. This property of QCD is supported by the evidence presented in \cite{ABFP18,AFP22,ADF23,AFIP23,FP24} and sets it apart from four-dimensional QED, putting it in the same class of theories as two-dimensional QED -- the Schwinger model \cite{Sch62a,Sch62b,LS71,Ito75,Nak76a}~--, which is known to be confining and massive. Second, this bound state generates massless poles in three- and four-point vertices of the theory, which are in turn transferred to the gluon polarization, evading the seagull identity which protects it from acquiring a mass. This is the so-called Schwinger mechanism, an excellent review of which in the context of QCD is given in \cite{FP25}. Third, the ``transfer'' of the poles is nothing more than the ``eating'' by the gluon of the bound state, which -- as we saw in this Letter -- ends up in the gluon spectrum. The latter arranges itself so that the bound state becomes the first parent state of the elementary BRST quartet -- hence the Goldstone boson associated to the spontaneous breaking of the Kugo-Ojima charge. Finally, once the color charge is corrected for symmetry breaking, the massiveness of gluons causes the vanishing of the surface integral of the chromoelectric field -- which completely determines the color of physical states --, leading to color confinement.

It is worth noting that, in this picture, BRST symmetry -- and the Ward-Takahashi (Slavnov-Taylor) identities of QCD with it -- are at all times preserved. Indeed, in the presence of gluon mass generation, the existence of the composite Goldstone boson is essential for the correct functioning of the quartet mechanism, which keeps the gauge degrees of freedom of QCD separated from its physical degrees of freedom.

\section{Acknowledgments}

We are grateful to A. Borys, C. Branchina, V. Branchina, D. Dudal, C. Gigliuto, G. La Rosa, J. Papavassiliou, M. Ruggieri, and F. Siringo for helpful discussions. This work was supported in part by PIACERI ``Linea di intervento 1'' (M@uRHIC) of the University of Catania and by PRIN2022 (Projects No. 2022SM5YAS and No. P2022Z4P4B) within Next Generation EU fundings.

\bibliography{QCD}

\end{document}